# Chapter: Exploring the Power of Creative AI Tools and Game-Based Methodologies for Interactive Web-Based Programming

Benjamin Kenwright

## Abstract

In recent years, the fields of artificial intelligence and web-based programming have seen tremendous advancements, enabling developers to create dynamic and interactive websites and applications. At the forefront of these advancements, creative AI tools and game-based methodologies have emerged as potent instruments, promising enhanced user experiences and increased engagement in educational environments. This chapter explores the potential of these tools and methodologies for interactive web-based programming, examining their benefits, limitations, and real-world applications. We examine the challenges and ethical considerations that arise when integrating these technologies into web development, such as privacy concerns and the potential for bias in AI-generated content. Through this exploration, we aim to provide insights into the exciting possibilities that creative AI tools and game-based methodologies offer for the future of web-based programming.

## 1 Introduction

In an era characterized by unprecedented technological advancement, the boundaries between the physical and digital realms have blurred, revolutionizing nearly every facet of human existence. Nowhere is this transformation more evident than in the domain of education, where the convergence of Creative AI Tools and Game-Based Methodologies with Interactive Web-Based Programming has ushered in a paradigm shift of monumental proportions (Kenwright, 2023, 2021b). The traditional classroom, once emblematic of static lectures and linear knowledge dissemination, is gradually giving way to a dynamic and interactive learning ecosystem that transcends the limitations of time and space. As we stand at the precipice of this educational renaissance, it becomes imperative to critically scrutinize the implications, possibilities, and challenges posed by this amalgamation of cutting-edge technologies.

This chapter serves as both a compass and a magnifying glass, guiding us through the intricate landscape shaped by the synergy of Creative AI Tools and Game-Based Methodologies within the realm of interactive web-based programming. The technological mosaic we encounter here holds the potential to redefine how knowledge is not just delivered, but absorbed, internalized, and applied. However, in our pursuit of the educational Holy Grail, we must tread carefully, for alongside the promises of innovation lay ethical quandaries that demand profound contemplation.

As we delve deeper into this transformative landscape, it becomes clear that Creative AI Tools are more than mere tools; they are the architects of personalized learning experiences that cater to the diverse needs and preferences of individual learners. Simultaneously, the infusion of Game-Based Methodologies injects an element of engagement and immersion, turning the educational journey into a narrative where learners are active participants, not passive recipients. Yet, we must be wary of the allure of novelty, as the rapid integration of technology can lead to a proliferation of solutions that are flashy but lack substance. The synthesis of these technologies also necessitates a reckoning with the ethical implications that arise. Issues surrounding data privacy, algorithmic biases, and the digital divide require our immediate attention. As the quest for personalized and immersive learning experiences advances, we must navigate the delicate balance between technological progress and safeguarding the rights and dignity of learners.

In navigating this uncharted territory, educators and learners themselves must become conscientious explorers. This chapter seeks to foster a nuanced understanding of the multifaceted landscape that unfolds when Creative AI Tools and Game-Based Methodologies intertwine with the canvas of interactive web-based programming. It invites us to ask probing questions, confront preconceived notions, and envisage the future of education that lies at the confluence of human ingenuity and technological prowess.

## 2 Advancements in AI and Web-Based Programming

### 2.1 Evolution of AI in Web Development

The Evolution of AI in Web Development stands as a testament to the relentless innovation that characterizes the digital age. From its humble beginnings as rudimentary chatbots to the sophisticated neural networks underpinning today's web applications, AI has woven itself into the very fabric of web development. Its evolution reflects a symbiotic relationship between human creativity and machine learning algorithms. As AI algorithms became more adept at understanding user preferences, web developers found themselves armed with tools to craft highly personalized and intuitive online experiences. This journey, however, has not been without its challenges. The pursuit of seamless human-machine interaction has led to ethical considerations, such as privacy infringements and algorithmic biases, casting a critical light on the rapid strides AI has taken in this domain. Thus, the evolution of AI in web development is a testament to the marvels of technological progress intertwined with the complex moral and societal questions that it brings to the fore.

### 2.2 Modern Trends in Web-Based Programming

Modern Trends in Web-Based Programming are a testament to the ever-evolving nature of the digital landscape. As the digital realm becomes more intertwined with our daily lives, web-based programming has transitioned from mere functionality to a dynamic conduit for expression and interaction (Kenwright, 2023). Responsive design has emerged as a cornerstone, ensuring seamless experiences across a multitude of devices and screen sizes. Moreover, the rise of Single Page Applications (SPAs) has redefined how content is presented, offering fluidity akin to native applications. This shift, however, has come with its share of challenges, with SPAs often posing concerns related to initial load times and search engine optimization. The advent of Progressive Web Apps (PWAs) addresses these issues, infusing web applications with the speed and offline capabilities of native apps. The symbiosis of web development and APIs has resulted in a thriving ecosystem of interconnected services, enabling developers to harness the power of external functionalities with ease. Yet, amidst the innovation lies a dichotomy; while these trends foster a more immersive and efficient web, they can inadvertently lead to a digital divide, where those without access to cutting-edge technologies risk exclusion. In dissecting these trends, it becomes evident that the contemporary web programmer is not merely a code craftsman but an orchestrator of user experiences, a guardian of accessibility, and a custodian of ethical considerations in an increasingly digital world.

Web 1.0 to 4.0 concepts that describe potential future stages of the internet's evolution beyond the current state of the Web 2.0. While these terms are not universally defined and can vary in their interpretations, they generally refer to advancements in technology and the way people interact with the online world. For example increasing dependency on automated technologes and machine learning for content and management of resources (see Figure 1).

## 3 Emergence of Creative AI Tools and Game-Based Methodologies

### 3.1 Convergence of Creativity and AI

The convergence of creativity and AI marks a pivotal moment in the evolution of both technological advancement and human artistic expression. The emergence of creative AI tools and game-based methodologies holds the promise of revolutionizing industries ranging from art and entertainment to education and problem-solving. However, it is imperative to approach this convergence with a nuanced and critical perspective, recognizing the intricate interplay between innovation and potential pitfalls.

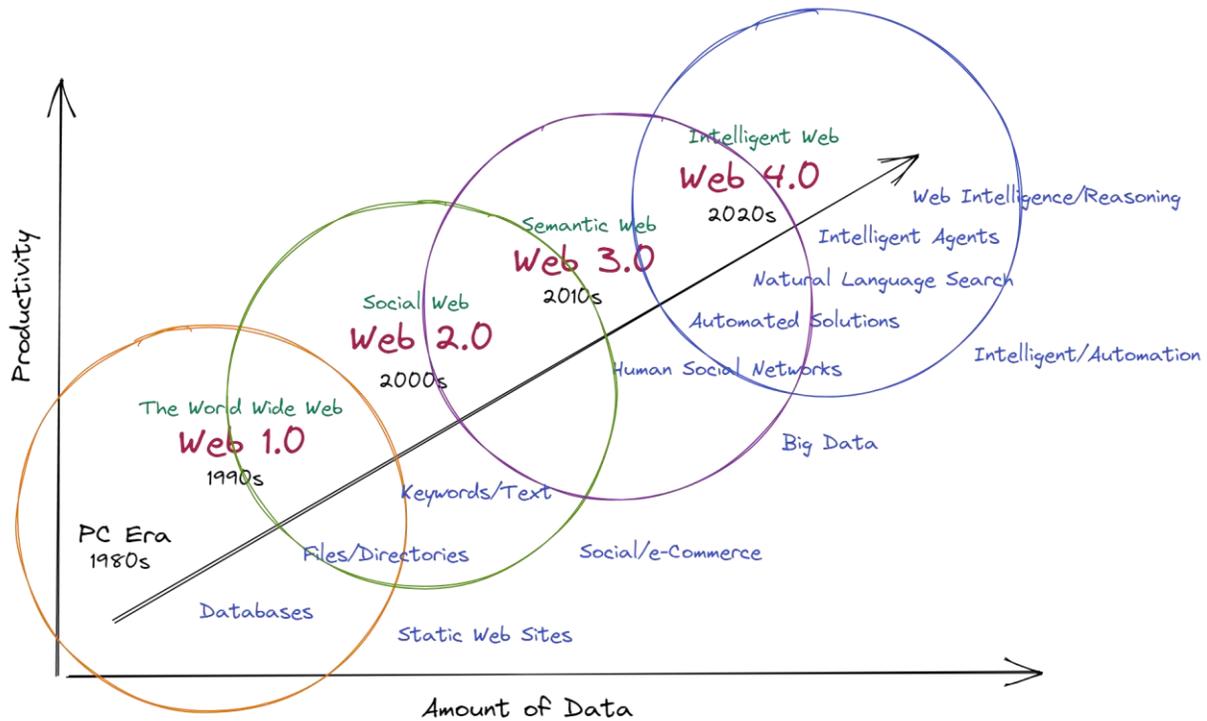

Figure 1: **Evolution of the Web** - Web 1.0 to Web 4.0

Creative AI tools have demonstrated remarkable capabilities in generating art, music, literature, and other creative content. They harness vast datasets and intricate algorithms to mimic human artistic processes, producing results that often blur the line between human-made and AI-generated creations. This raises fundamental questions about authorship, originality, and the value of human creativity. As AI systems learn from existing works, they risk perpetuating existing biases and artistic conventions, hindering the emergence of truly revolutionary and unconventional artistic forms. The sheer efficiency of AI-generated content could also contribute to an oversaturation of shallow, formulaic art, diluting the profound emotional and intellectual impact that genuine human creativity can evoke.

In the realm of game-based methodologies and education, the integration of AI presents a doubleedged sword. Gamification has shown remarkable potential in enhancing engagement, motivation, and learning outcomes. The utilization of AI to personalize learning experiences, offer real-time feedback, and adapt content to individual learners' needs can be transformative. However, a critical consideration lies in the potential devaluation of authentic learning experiences. The reliance on AI-driven metrics for success could overshadow the importance of fostering critical thinking, complex problem-solving, and collaborative skills that are vital in an ever-evolving world. The gamification of education, while enhancing motivation, must strike a careful balance to ensure that the pursuit of extrinsic rewards does not overshadow the intrinsic value of knowledge acquisition and holistic skill development.

Of course, the convergence of creativity and AI requires a discerning approach to ethics and transparency. The potential for AI-generated content to deceive audiences by masquerading as human-made creations raises ethical concerns about the authenticity of art and creative expression. The opacity of AI processes poses challenges in determining the origin and manipulation of content, impacting the credibility and trustworthiness of artistic endeavors. Additionally, the displacement of human creators in favor of AI-generated content could have profound socio-economic implications, altering employment landscapes and cultural paradigms.

## 3.2 Integration of Game Elements in Web Programming

The Integration of Game Elements in Web Programming represents a significant paradigm shift in the design and functionality of interactive online experiences. In response to evolving user expectations and technological capabilities, web developers have increasingly adopted principles from the realm of game design to enhance user engagement and interactivity. Elements such as gamification, interactive storytelling, and immersive interfaces are being seamlessly woven into the fabric of web applications, creating dynamic and captivating digital environments. This integration holds the potential to transform passive online interactions into active and participatory experiences, thereby fostering increased user retention and motivation. However, the successful incorporation of game elements into web programming demands a delicate balance between entertainment and utility. While the gamified approach can promote user interaction, it must be carefully tailored to align with the primary goals and content of the application, avoiding superficial engagement that detracts from the intended user experience (Kenwright, 2017, 2016b). As this trend continues to evolve, it prompts further exploration into how the principles of game design can be harnessed to optimize web-based applications for heightened engagement and usability, opening avenues for interdisciplinary research at the intersection of technology, psychology, and user-centered design.

# 4 Enhancing User Experience through Creative AI Tools

## 4.1 Personalization and Dynamic Content Generation

Personalization and Dynamic Content Generation stand as pivotal components within contemporary digital landscapes, reshaping user experiences and information dissemination (Figure 2). Enabled by sophisticated algorithms and data-driven insights, personalization tailors content delivery to individual preferences, enhancing relevance and engagement. Analyzing user behaviors and contextual cues, digital platforms curate content that aligns with users' interests, thus augmenting information accessibility and usability. Concurrently, dynamic content generation introduces an adaptive dimension to digital interactions. Utilizing real-time data and contextual variables, such systems dynamically craft content that responds to user inputs or environmental shifts. This fusion of personalization and dynamic content generation serves to optimize user engagement and interaction depth, driving prolonged interactions and fostering deeper connections between users and digital interfaces (Oh et al., 2018). However, this amalgamation also raises questions concerning privacy and ethical considerations (Miller et al., 2023). Striking the right balance between personalization and user privacy remains an ongoing challenge, necessitating nuanced approaches that respect user autonomy and consent. The interplay of these two facets exemplifies the evolving landscape of digital experiences, underscoring the interdisciplinary nature of research at the intersection of technology, user behavior, and ethics.

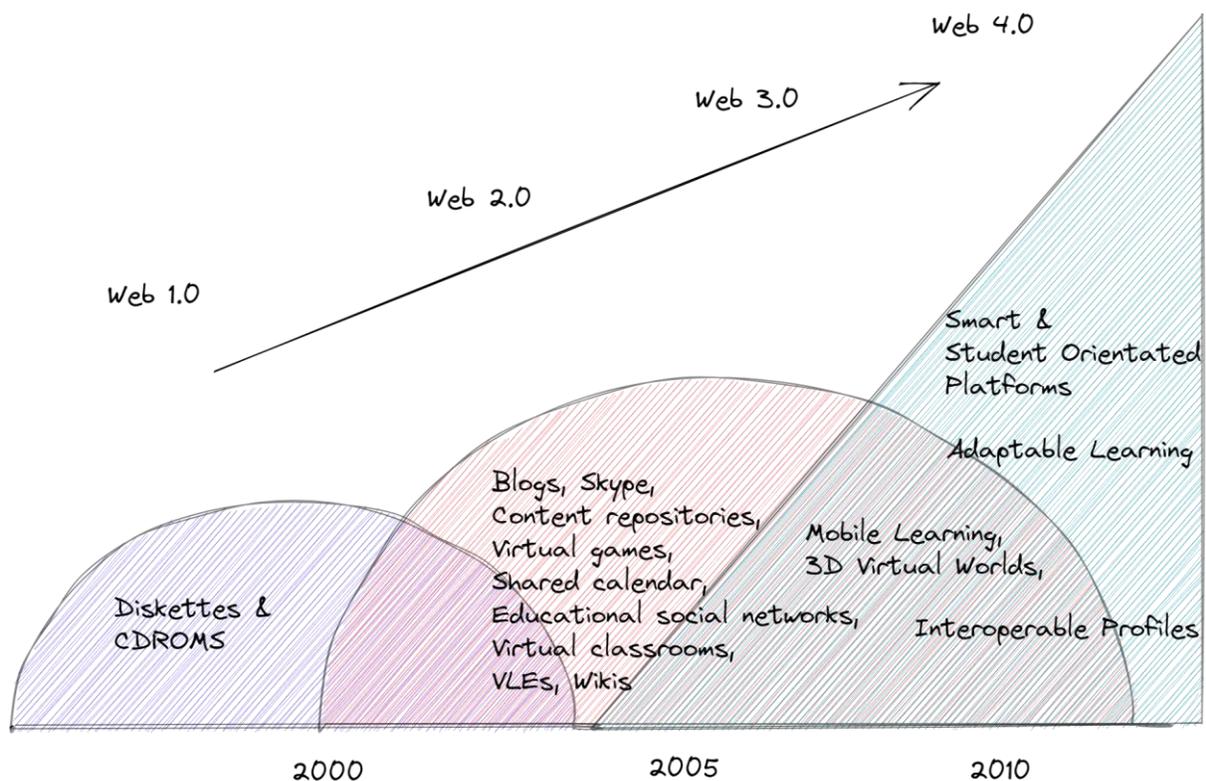

Figure 2: **Web Learning** - Evolution in terms of web-based concepts and technologies for learning.

### 4.2 Designing Intuitive User Interfaces with AI

Designing Intuitive User Interfaces with AI represents a profound transformation in the realm of user experience design, where technology becomes an active collaborator in the creation of interfaces that seamlessly adapt to human behaviors and preferences. Through AI-driven insights, interfaces can evolve from static arrangements of elements into dynamic ecosystems that anticipate user needs. Natural language processing and machine learning algorithms empower interfaces to understand user input contextually, responding with tailored suggestions and actions. However, this convergence also carries intrinsic challenges. The quest for intuition must not compromise user agency, as the overt presence of AI decision-making can potentially overshadow human control and comprehension. Furthermore, the ethical considerations surrounding AI-generated interfaces extend to transparency, bias mitigation, and data privacy. As the synergy between AI and design matures, the critical juncture lies in harmonizing technology's predictive capabilities with the user's desire for agency and transparency, thereby crafting interfaces that are not just intuitive but also respect human values and autonomy.

## 5 Engaging Web-Based Programming with Game-Based Approaches

### 5.1 Gamification for Enhanced Learning and Interaction

Gamification for Enhanced Learning and Interaction introduces a dynamic dimension to education and user engagement by infusing gameplay elements into traditionally instructional contexts. Rooted in the principles of game design, gamification leverages intrinsic motivators such as competition, achievement, and rewards to drive active participation and knowledge acquisition (Kenwright, 2020; Maulana et al., 2021). Transforming learning materials into interactive challenges and quests, gamification fosters a sense of accomplishment that resonates with learners' inherent desire for mastery. Moreover, the social components of gamified learning environments encourage collaboration and knowledge-sharing, cultivating a sense of community among participants. However, the pedagogical effectiveness of gamification is a complex interplay of design, content alignment, and user engagement. Striking the right balance between entertainment and educational objectives is paramount to ensure meaningful learning outcomes. As gamification reshapes learning paradigms, there is a growing call for research into its optimal implementation, tailored to diverse learning styles, age groups, and subject matters, thereby propelling the evolution of education into a more interactive and immersive realm (Figure 3).

### 5.2 Incorporating Interactive Storytelling in Web Development

Incorporating Interactive Storytelling in Web Development signifies a departure from passive content consumption towards a dynamic and immersive user experience (Sadik, 2008). Weaving narratives into the digital fabric, web developers have the potential to foster deeper connections and engagement. Interactive storytelling leverages multimedia elements, branching narratives, and user-driven choices to grant users agency over the narrative progression, blurring the lines between audience and author. However, this fusion presents multifaceted challenges. Striking the right balance between guided narratives and user autonomy is a delicate endeavor, as too much freedom can dilute the intended storyline, while excessive constraints might hinder the feeling of agency. Additionally, the technical complexities of creating seamless and responsive interactive narratives require meticulous planning and execution. Furthermore, ethical considerations emerge in cases where narratives touch upon sensitive topics or manipulate emotions to influence user behavior (Baldwin & Ching, 2017). As this narrative paradigm evolves, researchers and de-velopers are tasked with exploring the intersection of storytelling traditions, human psychology, and technological capabilities to craft immersive experiences that resonate both intellectually and emotionally with users.

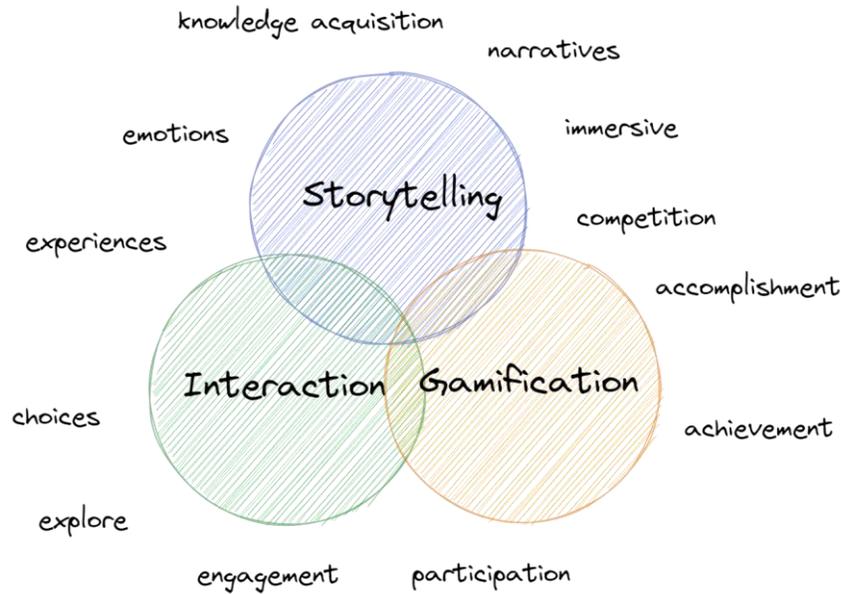

Figure 3: **Synergistic Benefits** - Combining gamification, interaction, and storytelling enhances learning by fostering engagement, active participation, and meaningful context that promotes effective knowledge acquisition and retention.

## 6 Unveiling the Potential: Benefits of Interactive Web-Based Programming

### 6.1 Amplifying User Engagement and Retention

Amplifying user engagement and retention represents a pivotal aspiration within contemporary digital environments, where the saturation of content and fleeting attention spans challenge the efficacy of online interactions. Leveraging a spectrum of techniques, from personalization to interactive elements, designers and developers endeavor to create interfaces that not only capture user attention but also encourage prolonged and recurring interactions. The integration of AI-driven recommendations tailors content delivery to individual preferences, enhancing the relevance of information presented. Moreover, the infusion of interactive components such as quizzes, polls, and user-generated content not only heightens engagement but also cultivates a sense of active participation (Kenwright, 2020). However, the attainment of sustained engagement involves intricate considerations. Designers must navigate the fine line between entertainment and utility, ensuring that the gamified elements or personalized content do not overshadow the primary purpose of the interface. Moreover, the ethical implications of data collection for personalization and the potential to inadvertently contribute to information bubbles necessitate thoughtful contemplation. As technology continues to redefine the digital landscape, the endeavor to amplify engagement and retention underscores the interdisciplinary nature of research, encompassing psychology, user behavior, design, and ethical considerations in shaping the trajectory of digital experiences.

## 6.2 Novel Approaches to Problem-Solving through Interactivity

Novel Approaches to Problem-Solving through Interactivity epitomize a transformative paradigm in the realm of cognitive engagement and solution discovery. The fusion of interactivity and problem-solving instigates a departure from conventional methods, allowing individuals to engage with challenges not as passive observers, but as active participants in the solution-finding process. Interactive problem-solving environments offer dynamic feedback loops, enabling users to experiment, iterate, and refine their strategies in real-time. This iterative nature of interactivity nurtures critical thinking, adaptability, and creativity, as users are encouraged to explore diverse avenues and adjust their approaches based on immediate outcomes. However, the successful implementation of interactive problem-solving hinges on the meticulous orchestration of interface design, user experience, and the cognitive load associated with decision-making. Overloading users with options can result in decision fatigue, hindering effective problem-solving (Fuad et al., 2018). Furthermore, the democratization of problem-solving, while empowering, also raises questions about the accuracy and quality of solutions generated through crowd-sourced approaches. This evolving landscape of interactive problem-solving necessitates a comprehensive understanding of human cognition, learning theories, and the role of technology in augmenting cognitive processes. Merging the realms of technology, psychology, and pedagogy, this paradigm invites critical inquiry into the optimization of interactivity as a catalyst for effective and innovative problem-solving in an array of domains (Hilliges et al., 2007).

# 7 Navigating Constraints: Limitations of Creative AI and GameBased Techniques

## 7.1 Technical Challenges and Feasibility in Educational Context

Technical Challenges and Feasibility in the educational context underscore a critical dimension of the integration of technology into learning environments. While the promise of technological tools for education is vast, their effective implementation is contingent on surmounting a range of technical hurdles. The diverse infrastructure and resources available to educational institutions, from K-12 to higher education, influence the feasibility of deploying advanced technologies uniformly. Disparities in internet connectivity, hardware availability, and software compatibility can impede equitable access to educational resources, inadvertently exacerbating educational inequalities. Moreover, the integration of complex technologies like virtual reality, artificial intelligence, or interactive simulations necessitates specialized expertise in both technological implementation and educational pedagogy. The potential for technical glitches, system failures, or data breaches further underscores the need for meticulous planning and robust technical support systems. Balancing the allure of cutting-edge tools with the pragmatic considerations of implementation and maintenance is imperative to ensure that technological interventions truly enhance the learning experience, promote inclusivity, and remain sustainable in diverse educational settings.

## 7.2 Addressing Cognitive Load and User Fatigue

Addressing cognitive load and user fatigue has become an increasingly pressing concern in the digital landscape, particularly in the context of creative AI and game-based techniques. While these approaches hold immense potential for enhancing user engagement and experiences, they also introduce a range of challenges that can exacerbate cognitive strain and fatigue. Creative AI, with its ability to generate vast amounts of content, runs the risk of overwhelming users with excessive choices and stimuli (Sevcenko et al., 2021). The sheer volume of AI-generated options, while providing variety, can lead to decision paralysis and cognitive overload. Furthermore, the uncanny valley effect, where AI-generated content closely mimics human creations but falls short in authenticity, can create a cognitive dissonance that detracts from the intended immersive experience.

Similarly, game-based techniques, despite their effectiveness in boosting motivation and learning, can inadvertently burden users with high cognitive demands. The integration of complex mechanics, intricate narratives, and competitive elements can elevate engagement but also increase cognitive load. Balancing the engagement factor with cognitive demands becomes pivotal to prevent users from feeling overwhelmed, frustrated, or mentally fatigued. Additionally, the overemphasis on extrinsic rewards and competitive benchmarks within gamified environments can lead to a cycle of stress and fatigue, undermining the intrinsic joy of learning and exploration (Ikehara et al., 2013).

Strategies to mitigate these challenges involve thoughtful design that aligns with human cognitive capacities. In the realm of creative AI, curating AI-generated content to provide a manageable number of options and focusing on enhancing the genuine quality rather than quantity can alleviate decision fatigue. Incorporating imperfections that highlight the AI's creative process rather than masking it can foster a sense of authenticity. In game-based methodologies, adopting adaptive learning mechanisms that tailor challenges to individual skill levels can prevent cognitive overload. Designing game mechanics that encourage collaboration, exploration, and intrinsic motivation over hyper-competition can alleviate the stress associated with excessive cognitive demands. The convergence of creative AI and game-based techniques offers both exciting opportunities and potential pitfalls in addressing cognitive load and user fatigue. As creators, designers, and developers navigate this landscape, they must remain vigilant in striking a delicate balance between innovation and cognitive well-being. Prioritizing human-centered design principles that reduce cognitive overload, prevent decision fatigue, and promote intrinsic motivation is crucial to ensure that these technologies ultimately enhance user experiences without compromising mental health and overall satisfaction.

# 8 Real-World Applications: Showcasing Innovative Implementations

## 8.1 AI-Driven Educational Platforms and Learning Portals

AI-driven educational platforms and learning portals stand as prominent exemplars of this potential, redefining the traditional boundaries of learning and instruction. Harnessing AI algorithms, these platforms can personalize learning journeys to cater to individual student needs, offering tailored content, adaptive assessments, and real-time feedback. The integration of machine learning can analyze vast amounts of data to identify learning patterns, allowing educators to make informed decisions about curriculum design and pedagogical approaches. Furthermore, AI-driven chatbots and virtual assistants provide on-demand support, enhancing the accessibility of learning materials and aiding in addressing student queries. These applications, however, are not immune to challenges. Ethical considerations concerning data privacy, algorithmic biases, and the potential replacement of human instructors must be navigated with prudence. As we witness the integration of AI into educational contexts, critical evaluation of its efficacy, user experiences, and ethical implications becomes essential in shaping the future of education and charting a path towards inclusive, personalized, and effective learning experiences (Kenwright, 2016b).

## 8.2 Game-Infused Learning Management Systems

Game-infused Learning Management Systems (LMS) represent a pioneering convergence of educational technology and game design principles within the realm of pedagogy. These systems leverage the intrinsic motivation and engagement inherent to games to enhance the educational experience. Incorporating gamified elements such as points, badges, leaderboards, and interactive challenges, LMS platforms transform learning into an immersive journey that encourages active participation and continuous progress. This approach resonates particularly with digital-native learners, capitalizing on their familiarity with gaming interfaces and mechanics. However, the integration of game elements within educational contexts demands a sophisticated balance between entertainment and educational objectives. Overemphasis on game mechanics can detract from the substance of learning, rendering the experience superficial. Furthermore, the efficacy of game-infused LMS hinges on clear alignment with learning outcomes, meaningful design of game mechanics, and ongoing assessment of their impact on learner engagement and knowledge retention (Kenwright, 2016a). As institutions increasingly embrace technology-enhanced education, the critical analysis of game-infused LMS is pivotal in discerning their potential to facilitate deeper understanding, cultivate critical thinking, and foster self-directed learning.

# 9 Ethical Considerations in Integrating AI and Game Elements

## 9.1 Privacy and Data Security Implications in Educational Settings

Privacy and data security implications in educational settings underscore a complex and evolving challenge brought about by the integration of technology into learning environments. While digital tools and platforms offer enhanced educational experiences, the collection, storage, and analysis of student data raise significant concerns regarding the protection of sensitive information. Educational institutions, entrusted with the stewardship of student data, must navigate a delicate balance between leveraging data for personalized learning and safeguarding the privacy rights of students and their families. The ubiquity of data breaches, coupled with the potential for misuse or unauthorized access, necessitates stringent protocols for encryption, access controls, and secure storage of educational data. Additionally, the ethical considerations of informed consent and transparent data usage become even more critical when dealing with minor students. Striking this balance requires collaboration among educators, administrators, policymakers, and technologists to implement robust data protection measures that not only ensure compliance with regulations but also foster a culture of trust and data responsibility within educational institutions.

## 9.2 Ensuring Transparency and User Consent for AI-Enhanced Learning

Ensuring transparency and user consent for AI-enhanced learning is an imperative at the intersection of technology and education. As artificial intelligence increasingly becomes a driving force in shaping educational experiences, the ethical obligation to maintain openness about its deployment is paramount. Learners, educators, and educational institutions must be provided with clear and comprehensible information about how AI algorithms are used to personalize content, assess performance, and make recommendations. Equally crucial is the principle of informed consent, where users understand and willingly agree to the data collection, analysis, and utilization that underlie AI-enhanced learning environments. Balancing the promise of enhanced educational outcomes with the preservation of user autonomy and data privacy requires educational technologists to design interfaces and mechanisms that empower users to make informed choices. Transparent communication not only cultivates trust but also empowers learners to actively engage with their own learning journey. In the era of AI, this critical juncture underscores the need for ethical reflection, legal compliance, and collaborative efforts to develop AI-enhanced educational environments that prioritize user agency, transparency, and responsible data practices.

# 10 Privacy Concerns in Interactive Web-Based Programming for Education

## 10.1 Balancing Data Collection and User Privacy

Balancing data collection and user privacy stands as a pivotal challenge in the digital age, where the seamless integration of technology into various facets of life has precipitated a wealth of personal data generation. In educational contexts, the collection of student data, fueled by the proliferation of digital learning platforms and AI-driven assessments, holds the promise of personalized learning experiences and informed instructional decisions. However, this potential must be meticulously weighed against the imperative of safeguarding user privacy. The extensive accumulation of sensitive information, ranging from academic performance to behavioral patterns, underscores the necessity for stringent data protection mechanisms. Striking this balance necessitates comprehensive data governance frameworks that emphasize informed consent, granular control over data sharing, and robust encryption protocols. Furthermore, ethical considerations extend to data anonymization, limiting the potential for identifying individuals through aggregation. The evolving regulatory landscape, exemplified by regulations like the General Data Protection Regulation (GDPR) and the Family Educational Rights and Privacy Act (FERPA), reflects the global recognition of the importance of striking this equilibrium. As educational institutions increasingly harness data for improved learning outcomes, the conscientious navigation of this equilibrium emerges as an interdisciplinary endeavor that amalgamates educational objectives, technological innovation, legal compliance, and ethical principles.

### 10.2 Safeguarding Student Information in AI-Enriched Learning Environments

Safeguarding student information in AI-enriched learning environments occupies a pivotal intersection of technological innovation, educational advancement, and ethical responsibility. As artificial intelligence takes on a larger role in educational settings, the vast array of data generated—ranging from academic performance to behavioral patterns—heightens concerns about data privacy and security. The amalgamation of sensitive information and AI analytics necessitates fortified measures to ensure that students' personal data remains confidential, unaltered, and inaccessible to unauthorized parties. The potential for unintended data breaches, algorithmic biases, and the creation of extensive digital profiles underscores the urgency of comprehensive data protection frameworks. However, this imperative must not hinder the potential of AI to transform education through personalized learning experiences and adaptive support mechanisms. Striking this balance requires a multifaceted approach encompassing robust encryption protocols, informed consent mechanisms, strict access controls, and ongoing audits. Moreover, it underscores the indispensable role of educators, administrators, policymakers, and technologists in cultivating a culture that prizes data privacy as a fundamental pillar of AI integration within educational contexts.

## 11 Bias Awareness: Navigating AI-Generated Content in Education

### 11.1 Identifying and Mitigating Bias in AI for Learning

Identifying and mitigating bias in AI for learning emerges as a critical endeavor in an era when machine learning algorithms wield significant influence in educational contexts. As AI algorithms inform decisions ranging from personalized content recommendations to assessment evaluations, the potential for bias—both latent and explicit—raises ethical concerns that demand systematic scrutiny. These biases can emanate from historical data, algorithm design, or even the inherent subjectivity of human-defined parameters. When left unchecked, biased AI systems perpetuate existing inequalities and disadvantage certain student groups, thereby undermining the pursuit of equitable education. The challenge lies in identifying and rectifying these biases through robust validation processes, continuous monitoring, and the deployment of fairness-enhancing techniques. Furthermore, transparency and interpretability in AI operations are paramount, enabling stakeholders to comprehend the decision-making process and unearth underlying biases. While AI has the potential to enhance educational outcomes, the critical imperative remains the ethical commitment to an unbiased, inclusive, and equitable learning environment, where technology serves as an ally in the pursuit of knowledge and not a perpetuator of discrimination.

**11.2 Ensuring Equity and Fairness in AI-Generated Educational Content**

Ensuring equity and fairness in AI-generated educational content is an intricate endeavor that requires a multifaceted approach, informed by the dual goals of harnessing the benefits of AI while safeguarding against the propagation of bias and inequality. As artificial intelligence plays an increasingly prominent role in shaping educational experiences, its potential to amplify or mitigate existing disparities becomes a focal point of ethical deliberation. AI algorithms, in their quest for efficiency and optimization, can inadvertently encode historical biases present in training data, thus perpetuating discriminatory outcomes in educational content, assessments, and recommendations. The recognition of this potential bias thrusts the responsibility onto educational technologists, developers, and content creators to proactively adopt strategies that identify, address, and mitigate bias throughout the AI development lifecycle.

Such strategies encompass not only the technical realm but also extend to the ethical, pedagogical, and societal dimensions. Leveraging diverse and representative datasets that encompass a spectrum of cultures, ethnicities, genders, and socioeconomic backgrounds is foundational in attenuating bias. This is accompanied by the ongoing refinement of algorithms to explicitly counteract bias, enabling AI systems to correct or modify results that deviate from equitable outcomes. Furthermore, transparency and interpretability of AI processes play a pivotal role in engendering trust among users and educational stakeholders. This transparency not only highlights potential sources of bias but also empowers educators and students to critically assess AI-generated content.

Pedagogically, an equitable AI-driven educational landscape necessitates the design of content that resonates with diverse learning styles and perspectives. The integration of culturally sensitive materials, the incorporation of multiple viewpoints, and the flexibility to adapt to individual learning trajectories are pivotal in fostering inclusive education. Additionally, educators are entrusted with the role of active mediators in the AI-infused classroom, offering context, critical analysis, and contextualization of AI-generated content to ensure that students develop the ability to discern, challenge, and counteract any potential biases.

On a broader societal level, equitable AI deployment demands collaboration among educators, researchers, policymakers, and the technology industry. This entails the creation of ethical guidelines and regulations that mandate fair and transparent AI practices in educational contexts, complemented by continuous assessment and audits to ensure compliance. Moreover, fostering technological literacy among students, educators, and parents is integral in enabling them to discern biased content, advocate for equitable AI practices, and contribute to the evolution of AI technologies that align with educational and societal values. Ensuring equity and fairness in AIgenerated educational content represents a multifaceted and evolving endeavor that engages the realms of technology, ethics, pedagogy, and societal norms. As AI becomes increasingly ingrained in the educational landscape, the responsibility to challenge, correct, and prevent bias rests with the collective efforts of educators, developers, policymakers, and learners themselves. Through a holistic and proactive approach, the educational community can harness the potential of AI to foster inclusive, transformative, and unbiased educational experiences that empower learners of diverse backgrounds and propel society toward a more equitable future.

# 12 The Future Landscape: Creative AI Tools and Game-Based Methodologies in Education

## 12.1 Evolving Trends in AI and Educational Technology

Evolving trends in AI and educational technology encapsulate a dynamic landscape that is redefining the fundamental paradigms of learning, instruction, and educational experiences. The integration of AI into education heralds a shift from traditional models of knowledge dissemination toward personalized, adaptive, and data-driven approaches that cater to individual student needs and preferences. Machine learning algorithms, empowered by large datasets, have the capacity to identify nuanced patterns in student performance, thereby enabling educators to tailor content, assessments, and interventions to optimize learning outcomes. The advent of AI-driven educational platforms augments the role of instructors as facilitators of knowledge, fostering a more learner-centric environment where students engage actively and collaboratively. Concurrently, virtual reality (VR) and augmented reality (AR) are breaking geographical constraints, enabling immersive and interactive learning experiences that span disciplines and contexts. However, as these trends burgeon, challenges loom large. Ethical concerns about data privacy, algorithmic biases, and the displacement of educators surface alongside the transformative potential of AI. The digital divide accentuates existing inequalities in access to advanced technologies, raising questions about the equitable distribution of AI-enhanced educational experiences. Furthermore, the human touch of education—the nuanced interaction, the mentorship, and the cultivation of critical thinking skills—must remain at the forefront even as AI advances, as it is these very aspects that shape informed, ethical, and empathetic citizens. Navigating these trends requires an interdisciplinary approach, engaging educators, technologists, policymakers, and ethicists in dialogue to ensure that AI-driven educational technologies serve not only as tools for efficiency but as conduits for meaningful, equitable, and holistic learning experiences that nurture the cognitive, emotional, and ethical dimensions of human development.

## 12.2 Envisioning the Pedagogical Impact of AI and Game Elements

As artificial intelligence and gamification strategies infiltrate educational landscapes, the traditional boundaries of pedagogy undergo a profound transformation. AI's predictive analytics and adaptive learning algorithms hold the potential to revolutionize education by tailoring instructional content and assessments to individual students' cognitive profiles and learning trajectories. This personalization aligns with constructivist principles, fostering a student-centered approach that accommodates diverse learning styles and paces. Moreover, AI's ability to provide immediate feedback and assistive interventions can cultivate metacognition, self-regulation, and lifelong learning skills. However, this shift necessitates a recalibration of the educator's role from knowledge transmitter to facilitator of critical thinking, fostering an environment where students actively engage with content rather than passively consume it. Simultaneously, the infusion of game elements into pedagogy taps into the innate human inclination towards play, creating immersive and interactive learning experiences that promote intrinsic motivation, collaboration, and problem-solving. Game-infused learning environments offer a fertile ground for situated cognition, where students learn by doing, exploring, and experiencing the consequences of their decisions in simulated scenarios. The gamified approach also aligns with social constructivist theories, encouraging collaborative learning and collective problem-solving within a competitive yet supportive framework. Yet, this transformation is not without challenges. The strategic alignment of game mechanics with learning objectives, the potential for superficial engagement, and the risk of extrinsic motivation overshadowing intrinsic curiosity demand meticulous design and pedagogical considerations. Moreover, ethical concerns arise in AI-driven systems, as data privacy, algorithmic bias, and the question of who controls the learning experience become paramount. In envisioning the pedagogical impact of AI and game elements, educators, researchers, and policymakers must collaborate to ensure that technology serves as a catalyst for enhanced learning experiences that cultivate critical thinking, foster collaboration, nurture curiosity, and empower learners with skills that transcend traditional disciplinary boundaries.

## 13 Case Study Example: Learning Success with Creative AI and Game-Based Techniques

This section delves into a real-world case study that exemplify the transformative potential of combining creative AI tools and game-based methodologies to enhance learning outcomes. These case studies showcase how innovative educational approaches have been implemented to create engaging, personalized, and effective learning experiences. Highlighting diverse instances of success, this section underscores the viability of integrating creative AI and game elements in educational settings.

### 13.1 Web Programming

This case study explores the successful integration of creative AI tools and game-based methodologies within a web programming course (Kenwright, 2020, 2021b), aimed at fostering interactive and engaging learning experiences. The course utilized web-based technologies such as WebGL, WebGPU, Web Neural Network API, and TensorFlow.js to create small game-based projects that served as both instructional tools and assessment measures.

The example demonstrates impact of combining creative AI and game elements in teaching web programming. Leveraging emerging web technologies, the aim was to enhance student engagement, promote active learning, and demonstrate the potential of these technologies in real-world web development contexts.

The course curriculum was designed around a series of small game-based projects that integrated web-based technologies like WebGL for rendering graphics, WebGPU for high-performance graphics processing, Web Neural Network API for machine learning integration, and TensorFlow.js for AI model deployment. Students were tasked with creating interactive web-based games that not only demonstrated technical proficiency but also showcased creative problem-solving and innovative design.

- Creative AI Integration: Students were introduced to the Web Neural Network API and TensorFlow.js (Smilkov et al., 2019), enabling them to incorporate AI-driven features within their game projects. For instance, students explored AI-powered opponent behavior, adaptive difficulty levels, and data-driven decisions to enhance gameplay experiences.
- Game-Based Learning: The use of various Web API, such as, WebXR and WebGPU (Kenwright, 2022, 2021a) facilitated the creation of visually appealing games, while game-based methodologies encouraged active learning and engagement. Students collaborated to develop multiplayer interactions, implement physics engines, and integrate interactive storytelling elements, which not only deepened their understanding of web programming but also allowed them to explore the nuances of user experience design.

During the course, the assess projects evaluated the students' technical skills, creativity, and ability to implement AI elements in their games. Through the iterative development process, students honed their coding abilities, learned to troubleshoot challenges, and gained proficiency in using AI-enhanced features.

The integration of creative AI and game-based techniques led to a significant increase in student engagement and enthusiasm. The practical application of web-based technologies in the context of game development motivated students to explore complex programming concepts in a tangible manner. The incorporation of AI not only added an innovative dimension to their projects but also exposed students to emerging technologies that have the potential to shape the future of web development.

## 13.2 Effective Learning Outcomes through Game-Based Instruction

The case study exemplifies the synergy between creative AI tools, game-based methodologies, and educational objectives. They provide tangible evidence of how these innovative approaches can transcend traditional instructional methods, fostering meaningful learning experiences that are engaging, personalized, and impactful. Showcases the successful fusion of creative AI tools and game-based methodologies in a web programming course, leveraging web-based technologies to create engaging and interactive learning experiences. The incorporation of WebGL, WebGPU, Web Neural Network API, and TensorFlow.js within small game-based projects not only enhanced students' technical skills but also exposed them to cutting-edge technologies that can shape the future of web development. This approach underscores the potential of innovative teaching methods to cultivate a generation of web developers who are not only proficient but also forwardthinking in their approach to technology.

# 14 Conclusion & Discussion

## 14.1 Envisioning the Future of Learning with Creative AI and Game-Based Approaches

The future of learning with creative AI and game-based approaches paints a compelling portrait of an educational landscape where innovation converges with pedagogy to redefine the boundaries of engagement and knowledge acquisition. The fusion of creative AI and game-based methodologies holds the promise of dismantling the conventional, one-size-fits-all model of education by ushering in a new era of personalized and immersive learning experiences. Creative AI's ability to generate dynamic content and adapt to individual learning patterns mirrors constructivist principles, facilitating active engagement, critical thinking, and problem-solving. Simultaneously, the integration of game-based elements taps into the intrinsic human drive for play, offering an environment where learners explore, experiment, and collaborate within a framework that melds competition and camaraderie. As students navigate AI-generated content within game-infused contexts, the boundaries between education and entertainment blur, creating an environment where learning becomes an organic and enjoyable pursuit. However, this vision is not immune to challenges. Ethical considerations surrounding data privacy, algorithmic biases, and the balance between human agency and AI-generated experiences pose significant questions. Educators, developers, and policymakers must collaboratively steer the trajectory of this evolution, ensuring that these approaches amplify pedagogical goals, cater to diverse learning styles, and cultivate skills that extend beyond mere content mastery. As we forge ahead, the future of learning stands as a testament to the harmonious coalescence of human ingenuity and technological prowess, illuminating a path toward holistic, dynamic, and learner-centric education.

## 14.2 Recap of Key Insights on Educational Transformation

The multifaceted exploration of themes such as AI-driven personalization, game-based methodologies, data privacy, and equity in education unveils the complex tapestry of opportunities and challenges within the ever changing educational landscape. The integration of creative AI tools has the potential to usher in personalized and adaptive learning experiences, aligning with constructivist principles and nurturing critical thinking skills. Simultaneously, the infusion of game elements capitalizes on intrinsic motivations, fostering engagement, collaboration, and problemsolving in educational contexts. However, the pursuit of these transformative approaches requires a vigilant consideration of ethical dimensions, including data privacy and algorithmic bias.

The overarching theme of equitable access to innovative learning experiences emerges as a central concern, with the digital divide underscoring the importance of addressing disparities in technology access. The collaborative effort of educators, technologists, policymakers, and researchers remains essential in harnessing the potential of educational transformation, guided by an ethos that prizes inclusivity, transparency, and learner agency. In reflection, these key insights coalesce to shape a comprehensive vision of education that embraces technological advancements while upholding the pedagogical values that nurture empowered, informed, and ethically responsible learners.

### 14.3 Prospects for the Convergence of AI, Games, and Web-Based Programming in Teaching and Learning

The prospects for the convergence of AI, games, and web-based programming in teaching and learning project a landscape of unprecedented educational innovation, interweaving technology, creativity, and pedagogy. This convergence holds the promise of reshaping educational paradigms by leveraging the strengths of each component. Creative AI tools can facilitate personalized learning experiences, adapting content and pace to suit individual needs, thereby catering to diverse learning styles and promoting deep understanding. Simultaneously, game-based methodologies introduce an element of engagement and interactivity, drawing from the intrinsic motivation associated with gameplay to drive student participation, collaboration, and critical thinking. Integrating these elements into web-based programming, educators can create dynamic, interactive, and responsive learning environments that transcend traditional classroom boundaries. However, realizing this potential necessitates navigating a complex array of considerations. Ethical concerns, including data privacy, algorithmic bias, and the responsible use of AI, must be addressed to ensure an equitable and inclusive learning experience. Furthermore, pedagogical expertise is crucial in orchestrating these convergent technologies to align with educational goals, avoiding superficial engagement and fostering deep, meaningful learning. The harmonious integration of AI, games, and web-based programming presents an exciting vista for teaching and learning, signifying not only a technological revolution but a pedagogical evolution that places the learner at the heart of an immersive, adaptive, and ethically informed educational journey.